\documentclass[11pt]{article}
\usepackage{}
\usepackage{amsfonts}
\usepackage{mathrsfs}
\usepackage{amssymb,amsmath,mathrsfs, amsthm}
\usepackage{palatino,cite}
\usepackage{graphicx}
\usepackage{hyperref,cases,anysize,enumerate}
\usepackage[usenames]{color}
\numberwithin{equation}{section}

\newtheorem{theorem}{Theorem}[section]

\newtheorem{corollary}{Corollary}[section]

\theoremstyle{definition}
\newtheorem{definition}{Definition}[section]

\theoremstyle{remark}

\date{}

\begin{document}

\title{{\bf{Geometric Hamiltonian matrix on the analogy between geodesic equation and Schr\"{o}dinger equation}}}
\author{ Jack Whongius\thanks{E-mail: fmsswangius@stu.xmu.edu.cn  }
\\
\vspace{.3 cm}
{\small School of Mathematical Sciences, Xiamen
University, China}\\
}

\maketitle

\begin{abstract}
By formally comparing the geodesic equation with the Schr\"{o}dinger equation on Riemannian manifold,  we come up with the geometric Hamiltonian matrix on Riemannian manifold based on the geospin matrix, and we discuss its eigenvalue equation as well. Meanwhile, we get the geometric Hamiltonian function only related to the scalar curvature. 

\end{abstract}


\section{Introduction and main results}
Let $x = (x^{ 1} ,\cdots,x^{n} )$ be Euclidean coordinates of ${{\mathbb{R}}^{n}}$, $M\subset {{\mathbb{R}}^{n}}$ open, ${{x}_{0}}\in M$. The
tangent space of $M$ at the point ${{x}_{0}}$,
${{T}_{{{x}_{0}}}}M$  is the space $\left\{ {{x}_{0}} \right\}\times B$, where $B$ is the $n-$dimensional vector space spanned by the basis
$\frac{\partial }{\partial {{x}^{1}}},\cdots ,\frac{\partial }{\partial {{x}^{n}}}$.  Here, $\frac{\partial }{\partial {{x}^{1}}},\cdots ,\frac{\partial }{\partial {{x}^{n}}}$  are the partial derivatives at the point ${x}_{0}$, hence,  $$v={{v}^{i}}\frac{\partial }{\partial {{x}^{i}}}\in {{T}_{{{x}_{0}}}}M$$ where ${{v}^{i}}=\frac{d{{x}^{i}}}{dt}={\dot{x}}^{i}$.   
Thus, let $x = (x^{ 1} ,\cdots,x^{n} )$ be local coordinates on manifold $M$. In these coordinates, a metric is represented by a positive definite, symmetric matrix ${{\left( {{g}_{ij}}\left( x \right) \right)}_{i,j=1,\cdots ,n}}$,   where the coefficients depend smoothly on $x$.  

Let $\left\langle \cdot ,\cdot  \right\rangle $ be the canonical real Euclidean inner product in the real $n-$dimensional linear space ${{\mathbb{R}}^{n}}$, $\left\| \cdot  \right\|$ be the norm induced by $\left\langle \cdot ,\cdot  \right\rangle $.  The product of two tangent vectors  $v={{v}^{i}}\frac{\partial }{\partial {{x}^{i}}},~c={{c}^{i}}\frac{\partial }{\partial {{x}^{i}}}\in {{T}_{p}}M$ at local point $p$ with coordinate representations $\left( {{v}^{1}},\cdots ,{{v}^{n}} \right)$ and $\left( {{c}^{1}},\cdots ,{{c}^{n}} \right)$,  then \cite{1} $$\left\langle v,c \right\rangle ={{g}_{ij}}\left( x\left( p \right) \right){{v}^{i}}{{c}^{j}},~~{{g}_{ij}}\left( x\left( p \right) \right)=\left\langle \frac{\partial }{\partial {{x}^{i}}},\frac{\partial }{\partial {{x}^{j}}} \right\rangle $$
Similarly, the length of $v$ is given by $\left\| v \right\|={{\left\langle v,v \right\rangle }^{1/2}}$. 

Let $x_{ 0} \in M$, and let $U$ be an open neighborhood of $x_{0}$ such that a chart for $M$ and a bundle chart for $B$ are defined on $U$. We thus obtain coordinate vector fields $\frac{\partial }{\partial {{x}^{1}}},\cdots ,\frac{\partial }{\partial {{x}^{n}}}$. For the Levi-Civita connection $\nabla$,  the so-called Christoffel symbols $\Gamma _{ij}^{k}\left( i,j,k=1,\cdots ,n \right)$ is then defined by \cite{2,3} $${{\nabla }_{{{\partial }_{i}}}}{{\partial }_{j}}=\Gamma _{ij}^{k}{{\partial }_{k}}$$where ${{\partial }_{j}}=\frac{\partial }{\partial {{x}^{j}}}$.    
A geodesic for a smooth curve $\gamma =\left[ a,b \right]\to M$, which satisfies in local coordinates \[{{\ddot{x}}^{i}}+\Gamma _{jk}^{i}{{\dot{x}}^{j}}{{\dot{x}}^{k}}=0,\text{ }i=1,\ldots ,n\]on $M$, where
\begin{equation}\label{a4}
  {\dot{x}^{j}}=\frac{d}{dt}{{x}^{j}}\left( \gamma \left( t \right) \right)={\dot{x}^{i}}\frac{\partial {{x}^{j}}}{\partial {{x}^{i}}}\left( \gamma \left( t \right) \right)={{v}^{i}}\frac{\partial {{x}^{j}}}{\partial {{x}^{i}}}\left( \gamma \left( t \right) \right)
\end{equation}
In physics, when the resultant force is zero on a object, the physical system is in equilibrium, and the object moves in geodesic motion. At this time, the acceleration on $M$ is expressed in the form of velocity
\begin{equation}\label{a1}
  {{\dot{v}}^{i}}+\Gamma _{jk}^{i}{{{v}}^{j}}{{{v}}^{k}}=0,\text{ }i=1,\ldots ,n
\end{equation}

In quantum mechanics, the most general form is the time-dependent Schr\"{o}dinger equation, which gives a description of a system evolving with time \cite{4}
 \begin{equation}\label{eq2}
   \sqrt{-1}\hbar \frac{\partial }{\partial t}\psi={{\hat{H}}}\psi
 \end{equation}
where $\partial/\partial t$ symbolizes a partial derivative with respect to time $t$, $\psi$  is the wave function of the quantum system. 
${{\hat{H}}}$
is the Hamiltonian operator in terms of the coordinates, where $V$ is the potential energy.

In order to see how the connection between the geodesic equation \eqref{a1} and Schr\"{o}dinger equation \eqref{eq2}, we rewrite the geodesic equation \eqref{a1} as 
\begin{equation}\label{a2}
  \sqrt{-1}\hbar\frac{d}{dt}{{v}^{i}}-\left(-\sqrt{-1}\hbar\Gamma _{jk}^{i}{{{v}}^{j}}{{{v}}^{k}} \right)=0,\text{ }i=1,\ldots ,n
\end{equation}
and Schr\"{o}dinger equation \eqref{eq2} is formally rewritten as 
\begin{equation}\label{a3}
  \left( \sqrt{-1}\hbar \frac{\partial }{\partial t}-{{\hat{H}}} \right) \psi=0
\end{equation}
As a result, such an analogy encourages us to consider intrinsic relation in form, it reveals that there exists Hamiltonian function induced by the geometric structure such that geometrodynamics can be described on the Riemannian manifold.  

The main result is organized as follows:
\begin{definition}
 The geometric Hamiltonian matrix on Riemannian manifold $M$ is given by 
  $$\hat{H}_{j}^{i}=-\sqrt{-1}\hbar W_{j}^{i}$$ where $W_{j}^{i}=\Gamma _{jk}^{i}{{v}^{k}}$ are the geospin variable.
\end{definition}
Note that the geometric Hamiltonian matrix on Riemannian manifold $M$ completely relies on the form of Christoffel symbols on coordinate chart. If $i=j$, then geometric Hamiltonian matrix turns to a energy function.  

\begin{theorem}\label{t1}
 For geospin variable $W_{j}^{i}=\Gamma _{jk}^{i}{{v}^{k}}$, then the geometric Hamiltonian function on Riemannian manifold $M$ is
  $H=-\sqrt{-1}\hbar {{w}^{\left( r \right)}}$ if $i=j$,  where ${{w}^{\left( r \right)}}=W_{j}^{j}$ is the geospin function and
  $H=\hat{H}_{j}^{j}$. 
\end{theorem}
Note that geospin function ${{w}^{\left( r \right)}}$ can be rewritten as ${{w}^{\left( r \right)}}={{A}_{k}}{{v}^{k}}$ in which
${{A}_{k}}={{\partial }_{k}}\ln \sqrt{\left| g \right|}$, where $g=\det \left( {{g}_{ij}}\left( x \right) \right)$. As for Riemannian volume form, 
any oriented pseudo-Riemannian (including Riemannian) manifold has a natural volume form. In local coordinates, it can be expressed as
$d\mu={\sqrt {|g|}}dx^{1}\wedge \dots \wedge dx^{n}$,
where the $dx^{i}$ are 1-forms that form a positively oriented basis for the cotangent bundle of the manifold. Here, $|g|$ is the absolute value of the determinant of the matrix representation of the metric tensor on the manifold.
\begin{corollary}\label{c1}
 For geospin function ${{w}^{\left( r \right)}}$ holds on Riemannian manifold $M$ with the Ricci flow, then the geometric Hamiltonian function is
  $H=\sqrt{-1}\hbar R$,  where $R$ is the scalar curvature.
\end{corollary}
Meanwhile, according to the Theorem \ref{t1}, the geodesic equation \eqref{a1} in a particular form follows  
$ \frac{d}{dt}{{v}^{i}}+{{w}^{\left( r \right)}} {{v}^{i}}=0$.   
Clearly, this obvious equation indicates the same mode as the Schr\"{o}dinger equation \eqref{a3}, we can collect two equations for a better understanding below, 
$$\begin{matrix}
   \left( \sqrt{-1}\hbar \frac{d}{dt}+\sqrt{-1}\hbar {{w}^{\left( r \right)}} \right){{v}^{i}}=0   \notag\\
   \left( \sqrt{-1}\hbar \frac{\partial }{\partial t}-\hat{H} \right)\psi^{j} =0  \notag
\end{matrix}$$where $\psi^{j}$ are the components of the wave function $\psi$.
This reveals the valid existence of the geometric Hamiltonian matrix on Riemannian manifold $M$.

\section{Geospin variables for geodesic equation}
In this section, we simply review the mathematical formalism involved in \cite{5},  it shows that the equation of geodesics can be simplified by geospin matrix, in the case of Riemannian and pseudo-Riemannian manifolds. In the theory of Riemann manifold and pseudo Riemann manifold, the expression of the coordinate space of the connection are Christoffel symbol, the connection matrix of connection is $\omega =\left( \omega _{i}^{k} \right)$ on the local coordinate $\left\{ {{x}^{i}} \right\}$,
where $\omega _{i}^{j}=\Gamma _{ik}^{j}d{{x}^{k}}$. The Levi-Civita connection is given by the Christoffel symbols
\begin{equation}\label{eq7}
  \Gamma _{ij}^{k}=\frac{1}{2}{{g}^{kl}}\left( \frac{\partial {{g}_{jl}}}{\partial {{x}^{i}}}+\frac{\partial {{g}_{il}}}{\partial {{x}^{j}}}-\frac{\partial {{g}_{ij}}}{\partial {{x}^{l}}} \right)
\end{equation}
where $g^{ ij}$ is the inverse of $g_{ ij}$, in particular, if $i=k$, then 
$\Gamma _{kj}^{k}={{A}_{j}}={{\partial }_{j}}\ln \sqrt{\left| g \right|}$ holds.

\begin{definition}\cite{5}\label{d1}
Let $(M,g)$ be a Riemannian manifold, two kinds of geospin variables can be defined as
\begin{equation}\label{eq3}
 W_{i}^{j}=\Gamma _{ik}^{j}{{v}^{k}},~~~{{W}_{ik}}=\Gamma _{ik}^{j}{{v}_{j}}
\end{equation}
where  ${{v}^{j}}, {{v}_{j}}$ are the components of  $v={{v}^{j}}\frac{\partial }{\partial {{x}^{j}}}={{v}_{j}}\frac{\partial }{\partial {{x}_{j}}}=\frac{dx}{dt}$ respectively.
\end{definition}
As geospin variables defined above, we mainly study the (1,1) form geospin variable $W_{i}^{j}$ that relates to the geospin matrix $W$ below.

\begin{definition}\label{d2}
The geospin matrix is defined as
$W=\left( W_{j}^{k} \right)$,
where $W_{j}^{k}$ are geospin variables.
\end{definition}
Obviously, the symmetric holds ${{W}_{kj}}={{W}_{jk}}$. The covariant derivative of a vector field $v$ can be rewritten as follows
\begin{equation}\label{eq4}
  {{\nabla }_{k}}{{v}^{j}}=\frac{\partial {{v}^{j}}}{\partial {{x}^{k}}}+W_{k}^{j},~~{{\nabla }_{k}}{{v}_{j}}=\frac{\partial {{v}_{j}}}{\partial {{x}^{k}}}-{{W}_{kj}}
\end{equation}
 As a consequence, the geodesic for Riemannian manifolds can be rewritten in a compact and simple form by using the geospin matrix as follows:
\begin{equation}
  \left\{ \begin{matrix}
   \frac{d{{x}}}{dt}={{v}}\begin{matrix}
   {} &  \\
\end{matrix}  \notag \\
   \frac{d{{v}}}{dt}=-Wv   \notag\\
\end{matrix} \right.
\end{equation}
This is a dynamic system. Consequently, the connection matrix of connection shows  $\omega =\left( W _{i}^{k} \right)dt$ on the local coordinate $x^{i}$,
where $\omega _{i}^{j}=W_{i}^{j}d{{t}}$ in which $d{{x}^{k}}={{v}^{k}}dt$ has been used. 

\section{Geometric Hamiltonian matrix on Riemannian manifold}
In this section, combining the physical meaning of the geospin matrix $W$ and comparison between geodesic equation and Schr\"{o}dinger equation, 
as a result, we define the geometric Hamiltonian matrix below.
\begin{definition}
 Based on geospin matrix $W$, then the geometric Hamiltonian matrix on $(M,g)$ is defined as
\[{{\hat{H}}^{\left( re \right)}}=-\sqrt{-1}\hbar W\]
The component is $\hat{H}_{j}^{i}=-\sqrt{-1}\hbar W_{j}^{i}$, where $W_{j}^{i}$ are geospin variables.
\end{definition}
The geospin matrix can be expressed in details as
\[W=\left( W _{i}^{j} \right)=\left( \begin{matrix}
    W _{1}^{1} & W _{2}^{1} & \cdots  & W _{n}^{1}  \\
   W _{1}^{2} &  W _{2}^{2} & \cdots  & W _{n}^{2}  \\
   \vdots  & \vdots  & \ddots  & \vdots   \\
   W _{1}^{n} & W _{2}^{n} & \cdots  &  W _{n}^{n}  \\
\end{matrix} \right)\]
where the matrix element of $W$ can be divided into two parts
\[W_{i}^{j}=\left\{ \begin{matrix}
   w^{(r)}=W _{i}^{i} ,i=j  \\
   W _{i}^{j},i\ne j  \\
\end{matrix} \right.\]
It is obvious that this directly leads to Theorem \ref{t1}.  
Obviously, the geometric Hamiltonian matrix on $(M,g)$ is an asymmetric matrix, so that its eigenvalues in generally are complex form. As a result, consider its eigenvalue equation \[{{\hat{H}}^{\left( re \right)}}X=-\sqrt{-1}\hbar WX={{\lambda }^{\left( re \right)}}X\]where $X$ represents the eigenvector in terms of the eigenvalue ${{\lambda }^{\left( re \right)}}$, and for geospin matrix $W$, its eigenvalue equation is assumed to be
\[WX=\left( {{\lambda }^{\left( s \right)}}+\sqrt{-1}{{\lambda }^{\left( im \right)}} \right)X=\lambda X\]where $\lambda= {{\lambda }^{\left( s \right)}}+\sqrt{-1}{{\lambda }^{\left( im \right)}}$ is the geometric frequency spectrum.
Thusly, then it yields
\begin{align}
-\sqrt{-1}\hbar WX& =-\sqrt{-1}\hbar \left( {{\lambda }^{\left( s \right)}}+\sqrt{-1}{{\lambda }^{\left( im \right)}} \right)X  \notag\\
 & =\left( \hbar {{\lambda }^{\left( im \right)}}-\sqrt{-1}\hbar {{\lambda }^{\left( s \right)}} \right)X  \notag\\
 & ={{\lambda }^{\left( re \right)}}X  \notag
\end{align}
then it results in a consequence
${{\lambda }^{\left( re \right)}}=\hbar {{\lambda }^{\left( im \right)}}-\sqrt{-1}\hbar {{\lambda }^{\left( s \right)}}$, these are the eigenvalues of the geometric Hamiltonian matrix ${{H}^{\left( re \right)}}$. Actually,  eigenvalues ${{\lambda }^{\left( re \right)}}$ form the complex energy spectrum of the geometric Hamiltonian matrix ${{H}^{\left( re \right)}}$.

In order to simplify the geospin matrix $W$ above, we consider $1\times 1$ $W$-matrix, namely, $W_{1}^{1}={{w}^{\left( r \right)}}$, then its eigenvalue
$\lambda ={{\lambda }^{\left( s \right)}}={{w}^{\left( r \right)}}$, the only one eigenvalue of geometric Hamiltonian matrix follows ${{\lambda }^{\left( re \right)}}=-\sqrt{-1}\hbar {{w}^{\left( r\right)}}$.

For a better analysis, we denote
\[{{W}^{\left( r \right)}}=\left( \begin{matrix}
    W _{1}^{1} & 0 & \cdots  & 0  \\
   0 &  W _{2}^{2} & \cdots  & 0  \\
   \vdots  & \vdots  & \ddots  & \vdots   \\
   0 & 0 & \cdots  &  W _{n}^{n}  \\
\end{matrix} \right),~~{{W}^{\left( a \right)}}=\left( \begin{matrix}
   0 & W_{2}^{1} & \cdots  & W_{n}^{1}  \\
   W_{1}^{2} & 0 & \cdots  & W_{n}^{2}  \\
   \vdots  & \vdots  & \ddots  & \vdots   \\
   W_{1}^{n} & W_{2}^{n} & \cdots  & 0  \\
\end{matrix} \right)\]
then the geospin matrix $W$ can be rewritten as  $W=\left( W_{i}^{j} \right)={{W}^{\left( r\right)}}+{{W}^{\left( a \right)}}$, as a result,  the geometric Hamiltonian matrix rewrites
\[{{\hat{H}}^{\left( re \right)}}=-\sqrt{-1}\hbar W=-\sqrt{-1}\hbar {{W}^{\left( r \right)}}-\sqrt{-1}\hbar {{W}^{\left( a \right)}}\]
We mainly focus on ${{H}^{\left( r \right)}}=-\sqrt{-1}\hbar {{W}^{\left( r \right)}}$ individually, more precisely, \[{{H}^{\left( r \right)}}=-\sqrt{-1}\hbar {{W}^{\left( r \right)}}=-\sqrt{-1}\left( \begin{matrix}
   {{E}^{\left( 1 \right)}} & 0 & \cdots  & 0  \\
   0 & {{E}^{\left( 2 \right)}} & \cdots  & 0  \\
   \vdots  & \vdots  & \ddots  & \vdots   \\
   0 & 0 & \cdots  & {{E}^{\left( n \right)}}  \\
\end{matrix} \right)\]
where ${{E}^{\left( 1 \right)}}=\hbar  W _{1}^{1}$, accordingly, it gets  ${{H}^{\left( 1 \right)}}=-\sqrt{-1}{{E}^{\left( 1 \right)}}=-\sqrt{-1}\hbar  W _{1}^{1}$, as a result of this formula, ${{H}^{\left( r \right)}}=\text{diag}\left( {{H}^{\left( 1 \right)}},\cdots ,{{H}^{\left( n \right)}} \right)$ is given accordingly. 

For continuum mechanics, the corresponding interpretation of strain can be made for Ricci flow, the Ricci flow equation \cite{6,7} is the evolution equation $d{{g}_{im}}(t)/dt=-2{{R}_{im}}$ for a Riemannian metric ${{g}_{ij}}$, where $R_{ij}$ is the Ricci curvature tensor. Hamilton (1982) showed that there is a unique solution to this equation for an arbitrary smooth metric on a closed manifold over a sufficiently short time. The concept of Ricci flow $\left( M,{{g}_{im}}\left( t \right) \right)$ opens a door for the combination of rational mechanics and modern physics. 

In geometry, the change of curvature can be explained by local internal rotation. The mechanical geometric explanation of Ricci flow is that the intrinsic curvature change is the reason for the metric change of closed manifold. Thus, the local internal rotation is attributed to the internal cause of the geometric evolution of the closed manifold configuration. If the internal rotation is not zero, the closed manifold will evolve until it reaches an equilibrium configuration. The Ricci flow is a powerful technique that integrates geometry, topology, and analysis. The idea is to set up a PDE that evolves a metric according to its Ricci curvature.
\subsection{Proof of the corollary}
\begin{proof}
  According to the expression of the geospin function
${{w}^{\left( r\right)}}=W_{i}^{i}={{A}_{j}}{{v}^{j}}$,
where $\Gamma _{ij}^{i}={{A}_{j}}=\frac{\partial \ln \sqrt{g}}{\partial {{x}^{j}}}$ has been used and here $g>0$. What's more, according to \eqref{a4}, it has
\begin{equation}\label{a5}
  \frac{d\ln \sqrt{g}}{dt}={{w}^{\left( r \right)}}={{v}^{j}}\frac{\partial \ln \sqrt{g}}{\partial {{x}^{j}}}
\end{equation}
By using the formula ${{\Gamma }^{i}}_{ki}=\frac{1}{2}{{g}^{im}}\frac{\partial {{g}_{im}}}{\partial {{x}^{k}}}=\frac{\partial \log \sqrt{g}}{\partial {{x}^{k}}}$,
plugging it to \eqref{a5}, it yields
\begin{equation}\label{a6}
  {{w}^{\left( r \right)}}=\frac{1}{2}{{g}^{im}}\frac{\partial {{g}_{im}}}{\partial {{x}^{j}}}{{v}^{j}}=\frac{1}{2}{{g}^{im}}\frac{d{{g}_{im}}}{dt}
\end{equation}
Hence, by substituting Ricci flow for \eqref{a6}, it leads to the Ricci scalar  ${{w}^{\left( r \right)}}=-{{g}^{im}}{{R}_{im}}=-R$, hence, the Theorem \ref{t1} can deduce the previously corollary \ref{c1}.

\end{proof}
Note that the geometric Hamiltonian function is a pure imaginary number which is only a function in terms of the scalar curvature $R$.

\end{document}